\documentclass[3p,10pt]{elsarticle}

\usepackage{amssymb}
\usepackage{amsmath}
\usepackage{amsthm}
\usepackage{amsfonts}
\usepackage{graphicx,epstopdf}
\usepackage{dcolumn}
\usepackage{bm}
\usepackage{mathtools}
\usepackage{hyperref}

\journal{Physical Letters A}

\begin{document}

\begin{frontmatter}

\title{Complexity of the Laughlin wave function from the Dyson-orbital perspective}
\author[1,2]{J. M. Zhang}
\ead{wdlang06@163.com}
\author[3]{Y. Liu}
\ead{liu\_yu@iapcm.ac.cn}
 \affiliation[1]{organization={Fujian Provincial Key Laboratory of Quantum Manipulation and New Energy Materials,
College of Physics and Energy, Fujian Normal University},
           % addressline={Fujian Normal University},
            city={Fuzhou},
        %    postcode={},
         %   state={},
            country={China}}
\affiliation[2]{organization={Fujian Provincial Collaborative Innovation Center for Advanced High-Field
Superconducting Materials and Engineering},
      %   addressline={},
          city={Fuzhou},
        %    postcode={},
         %   state={},
            country={China}}
\affiliation[3]{organization={Laboratory of Computational Physics, Institute of Applied Physics and Computational Mathematics},%Department and Organization
        %  addressline={},
            city={Beijing},
         %   postcode={100088},
      %     state={},
            country={China}}

\begin{abstract}
The Fermi sea is a simple and common concept in physics. However, a related and equally simple concept---the Dyson orbital---is far less discussed in physics, especially in textbooks. 
Yet, Dyson orbitals offer a valuable tool for characterizing the complexity of a fermionic wave functions, particularly in distinguishing between Fermi-sea-like and non-Fermi-sea-like states. As a preliminary application, we examine the Laughlin wave function and find the fortunate fact that the Dyson orbitals can be determined analytically. Further numerical data provides \emph{quantitative} evidence that the Laughlin wave function describes a strongly correlated, non-Fermi liquid state. 
\end{abstract}



\end{frontmatter}


\section{Introduction}
The Fermi sea is a fundamental concept in physics. As the ground state of a noninteracting Fermi gas, it is introduced in undergraduate courses of quantum mechanics, statistical mechanics, solid state physics, nuclear physics, etc. Not only is it simple theoretically,  but it also proves useful and relevant for real systems. For closed-shell atoms or nuclei, the ground state of the system often resembles a Fermi sea. In solid state physics, in the Sommerfeld theory of metals, the mere assumption that the ground state of the electrons is a Fermi sea alone can already account for many characteristics of metals \cite{mermin, giuliani, haldane}.
Recently, a Fermi sea consisting of noninteracting ultracold fermionic atoms has been realized experimentally \cite{esslinger}, with the characteristic Fermi surface demonstrated in a vivid way.

The Fermi sea is featured by its wave function---a Slater determinant---which represents the simplest form of a fermionic wave function. 
However, not all fermionic wave functions are Slater determinants or even closely resemble them. A notable example is the famous Laughlin wave function \cite{laughlin}, which captures the essential physics of the fractional quantum Hall systems. 

An interesting problem is how to characterize the complexity of a fermionic wave function \cite{ojanen} which cannot be put in the Slater determinant form. In this paper, we would like to promote the notion of Dyson orbital \cite{ortiz, goscinski,  pickup,linderberg} for this purpose, both in teaching and in research. This notion arises naturally  if we take the perspective of how  a fermionic wave function builds up particle by particle. As we shall see, it is a simple notion readily accessible to undergraduate and graduate students. Moreover, the Dyson orbitals associated with the Laughlin wave function can be determined analytically, which greatly facilitates the study of the latter's complexity. The upshot is that the Laughlin wave function is proven quantitatively to be a non-Fermi liquid state. 

\section{Dyson orbital}

The picture of a Fermi sea is that the single-particle levels are filled one by one. This picture is now even fulfilled experimentally with cold atoms \cite{joachim}. Let the eigenstates of the single-particle Hamiltonian, ordered with increasing energy,  be $\phi_n $ ($1 \leq n \leq \infty $). The $N$-particle ground state, which is a Fermi sea, is the state with the lowest $N$ levels filled. Explicitly, 
\begin{eqnarray}\label{fs1}
  |GS_N\rangle  = \prod_{n=1}^{N } \hat{a}^\dagger_{\phi_n } |vac \rangle . 
\end{eqnarray}
Here $\hat{a}_{\phi_n}^\dagger $ is the creation operator of single-particle orbital $\phi_n $. We have between two successive ground states, 
\begin{eqnarray}\label{fs2}
  |GS_{N+1} \rangle  = \hat{a}_{\phi_{N+1}}^\dagger |GS_N\rangle , 
\end{eqnarray}
or 
\begin{eqnarray}\label{fs3}
  |\langle GS_{N+1} | \hat{a}_{\phi_{N+1}}^\dagger |GS_N\rangle |^2 = 1 . 
\end{eqnarray}
That is, the $(N+1)$-particle Fermi sea is constructed by putting an extra particle on top of the $N$-particle Fermi sea. The point is that the newly added particle does not disturb the already present particles. 

In the presence of interaction, things are more complicated. First, the ground state of an $N$-particle system is generally not a Fermi sea, or it generally cannot be written in the form of (\ref{fs1}), where now the orbitals $\{ \phi_n \}$ are to be understood as some orthonormal basis (unnecessarily the eigenstates of the single-particle Hamiltonian). Second, the $(N+1)$-particle ground state does not relate to the $N$-particle ground state in the simple way as in (\ref{fs2}) or (\ref{fs3}). The $N$ particles generally will \emph{reorganize} themselves in response to the introduction of the $(N+1)$th particle. 

The first point prompts the problem of the optimal Slater approximation of a generic fermionc wave function, which was studied in Refs.~\cite{aoto,zhang1, zhang2, zhangnjp}. The second point prompts yet another optimization problem. For given $(N+1)$-particle state $|GS_{N+1} \rangle$ and $N$-particle state $|GS_N\rangle $, which (normalized to unity) single-particle orbital $f$ will maximize the quantity \cite{phe}
\begin{eqnarray}\label{defo}
  \mathcal{O}(f) = |\langle GS_{N+1} | \hat{a}_{f}^\dagger |GS_N\rangle |^2,
\end{eqnarray}
which is understood as the overlap between the two $(N+1)$-particle states $\hat{a}_f^\dagger |GS_N\rangle $ and $|GS_{N+1} \rangle$? Note that $\mathcal{O}$ can also be interpreted the other way round, namely, as the overlap between the two $N$-particle states $\hat{a}_f  |GS_{N+1}\rangle $ and $|GS_{N} \rangle$. The maximal value of $\mathcal{O}$ will be denoted as 
\begin{eqnarray}\label{omax}
  \mathcal{O}_{max} = \max_{\|f \|=1} \mathcal{O}(f) = \max_{\|f \|=1} |\langle GS_{N+1} | \hat{a}_{f}^\dagger |GS_N\rangle |^2.
\end{eqnarray}

It is easy to show that this quantity is upper bounded by unity (if both $|GS_N\rangle$ and $|GS_{N+1}\rangle$ are normalized to unity), 
\begin{eqnarray}\label{upperb}
  \mathcal{O}(f) &=& |\langle GS_{N+1} | \hat{a}_{f}^\dagger |GS_N\rangle |^2  \nonumber \\
   &\leq & \langle GS_{N} | \hat{a}_f \hat{a}_{f}^\dagger |GS_N\rangle  \nonumber  \\
   &=&  \langle GS_{N} | 1-  \hat{a}_{f}^\dagger \hat{a}_f |GS_N\rangle   \nonumber  \\
   &\leq &  \langle GS_{N}  |GS_N\rangle  = 1 . 
\end{eqnarray}
The first inequality is a Cauchy-Schwarz inequality---the matrix element of $\hat{a}_f^\dagger $ is interpreted as the overlap between the state $|GS_{N+1}\rangle $, which is normalized, and the state $\hat{a}_{f}^\dagger |GS_N\rangle $. From the second inequality we see that a necessary condition for $\mathcal{O}_{max}$ to achieve the upper bound is that the orbital $f$ is not occupied in $|GS_N \rangle $. 

We have a simple interpretation of the functional $\mathcal{O }(f) $. As defined in (\ref{defo}), it makes sense for an arbitrary single-particle orbital $f$, normalized or not.  Apparently, for given $|GS_N\rangle $ and $|GS_{N+1}\rangle $, the matrix element $\langle GS_{N+1} | \hat{a}_{f}^\dagger |GS_N\rangle $ is linear in $f$. Therefore, by the Riesz representation lemma, it is the inner product between $f$ and some single-particle orbital. That is, 
\begin{eqnarray}\label{matelem}
  \langle GS_{N+1} | \hat{a}_{f}^\dagger |GS_N\rangle = \langle \tilde{\phi}_D | f \rangle \equiv   \int dx \tilde{\phi}_{D}^*(x) f(x),
\end{eqnarray}
for some orbital $\tilde{\phi}_D$. As the subscript and the tilde indicate, this orbital is the so-called unnormalized Dyson orbital. The normalized Dyson orbital will be denoted as $\phi_D$.  
Plugging in (\ref{matelem}) the expression of the operator $\hat{a}_f^\dagger $, 
\begin{eqnarray}
  \hat{a}_f^\dagger  =\int dx  f(x) \hat{\psi}^\dagger(x) ,
\end{eqnarray}
we get
\begin{eqnarray}\label{dyson}
  \tilde{\phi}_{D}(x) = \langle GS_{N}|\hat{\psi}(x)|GS_{N+1}\rangle . 
\end{eqnarray}
Therefore,  $\mathcal{O}(f)$ is just the squared modulus of the overlap between $\tilde{\phi}_{D} $ and $f$,
\begin{eqnarray}\label{oexp}
   \mathcal{O} (f)  = \left |\int dx \tilde{\phi}_{D}^*(x) f(x)  \right |^2 = |\langle  \tilde{\phi}_{D} | f\rangle  |^2.
\end{eqnarray}
By the Cauchy-Schwarz inequality, we have then (note that $f$ is normalized)
\begin{eqnarray}\label{upperb3}
  \mathcal{O} (f) \leq  \int dx |f (x)  |^2  \int dx |\tilde{\phi}_{D}(x)  |^2 =  \int dx |\tilde{\phi}_{D}(x)  |^2 .
\end{eqnarray}
The equality is achieved when and only when $f$ is proportional to $\tilde{\phi}_{D}  $,  or when $f$ is essentially $\phi_D$. The significance of the Dyson orbital is then that it maximizes the functional $\mathcal{O}(f)$ in (\ref{defo}). The maximal value $\mathcal{O}_{max} = \| \tilde{\phi}_D  \|^2$. The inequality of (\ref{upperb}) implies that $\| \tilde{\phi}_D  \|\leq 1 $.  By (\ref{oexp}), we know that if $f_i $ ($1\leq i \leq \infty $) are an orthonormal single-particle basis, 
\begin{eqnarray}\label{sumrule}
  \sum_{i=1}^\infty  \mathcal{O}(f_i ) =\sum_{i=1}^{\infty} |\langle f_i | \tilde{\phi}_{D}\rangle |^2 =    \| \tilde{\phi}_D  \|^2  = \mathcal{O}_{max}. 
\end{eqnarray}

Let the wave functions of $|GS_N \rangle $ and $|GS_{N+1}\rangle $ in the first quantization formalism be $\Psi_{N}(x_1, \ldots, x_N )$ and $\Psi_{N+1}(x_1, \ldots, x_{N+1})$, respectively. We have then 
\begin{eqnarray}
  |GS_{N}\rangle  &=& \frac{1}{\sqrt{N!}}\int d x_1 \ldots d x_N \Psi_N(x_1, \ldots, x_N ) \hat{\psi}^\dagger(x_1)\ldots \hat{\psi}^\dagger(x_N)|vac\rangle ,  \\
  |GS_{N+1}\rangle  &=& \frac{1}{\sqrt{(N+1)!}}\int d x_1 \ldots d x_{N+1} \Psi_{N+1}(x_1, \ldots, x_{N+1} )  \hat{\psi}^\dagger(x_1)\ldots \hat{\psi}^\dagger(x_{N+1})|vac\rangle. 
\end{eqnarray}
Substituting these expressions into (\ref{dyson}), we get 
\begin{eqnarray}\label{dyson233}
  \tilde{\phi}_{D}(x) &=&\frac{1}{N!\sqrt{N+1}}\int d x_1 \ldots d x_N d y_1  \ldots d y_{N+1} \Psi_N^*(x_1, \ldots, x_N ) \Psi_{N+1}(y_1, \ldots, y_{N+1} ) \nonumber\\
  & & \quad \quad \quad \quad \quad \quad \quad \quad  \quad \times   \langle vac | \hat{\psi}(x_N)\ldots \hat{\psi}(x_1) \hat{\psi}(x) \hat{\psi}^\dagger(y_1)\ldots \hat{\psi}^\dagger(y_{N+1})|vac\rangle .
\end{eqnarray}
Here we note that the matrix element vanishes unless the set $\{x, x_1, \ldots, x_N\}$ is equal to the set $\{y_1, \ldots, y_{N+1}\}$, and the wave function $\Psi_{N+1}(y_1, \ldots, y_{N+1} ) $ and the operator product $\hat{\psi}^\dagger(y_1)\ldots \hat{\psi}^\dagger(y_{N+1}) $ are both antisymmetric under permutations of the $y$'s. This latter fact contributes a factor $(N+1)!$ to the integral. In the end,  we get an explicit expression of the \emph{unnormalized} Dyson orbital
\begin{eqnarray}\label{dyson2}
  \tilde{\phi}_{D}(x) =   \sqrt{N+1}\int dx_1 \ldots d x_{N} \Psi_N^*(x_1, \ldots, x_N) \Psi_{N+1} (x, x_1, \ldots, x_N ).  
\end{eqnarray}
If $|GS_{N}\rangle $ and $|GS_{N+1 }\rangle $ are Fermi seas defined as in (\ref{fs1}), either with this Equation or Equation (\ref{dyson}), we can obtain the result that $\tilde{\phi}_D$ is equal to $\phi_{N+1}$. 

A few remarks are in order. Here we introduce the Dyson orbital by examining the structure of fermionic wave functions without reference to any Hamiltonian or any physical process.
It is worth noting that this is not how this concept was introduced into physics. The concept is attributed to Dyson, likely because of his foundational work in the theory of Green functions \cite{dysonpaper}. Indeed, explicit expressions like (\ref{dyson}) emerge naturally in the Lehmann spectral representation of the Green functions \cite{fetter}. Nevertheless, identifying them as single-particle orbitals and serious study of their properties began only  much later in quantum chemistry, specifically, through the study of electron ionization and capture processes \cite{pickup,linderberg}. It is no wonder, as these processes are processes in which the electron number changes by one.  In these circumstances, under the so-called sudden approximation, the Dyson orbital is involved in the expressions of the most concerned transition intensities \cite{manne}. 
More recently, Dyson orbitals are also found mandatory to rationalize scanning tunnelling spectroscopy measurements of strongly correlated molecules \cite{jacs}. These are again processes with the electron number changing by one. Besides that, the Dyson orbital, or more precisely, the quantity $\mathcal{O}$ in (\ref{defo}), also finds use in nuclear physics, where it is called the ``spectroscopic factor'' and is used as a measure of the purity of the single-particle state $f$ \cite{ring, magic}. In contrast to its established presence in quantum chemistry and nuclear physics, the Dyson orbital's application in condensed matter physics is negligible, if not zero, as far as we know. 
%
% However, in quantum chemistry and nuclear physics, where the Dyson orbital is most discussed, it is traditionally introduced in the study of particle attachment or detachment processes \cite{ortiz, pickup, linderberg,ring}. Actually, the form of $\mathcal{O}$ in (\ref{defo}) suggests that it can be proportional to the squared modulus of some matrix element of some physical observable. In nuclear physics, $\mathcal{O}$ is called the ``spectroscopic factor'' and is used as a measure of the purity of the single-particle state $f$ \cite{ring, magic}. 

For readers familiar with the Fermi liquid theory, Equation (\ref{defo}) should evoke the spectral function. By definition (Ref.~\cite{giuliani}, page 446), a Fermi liquid state has the property that 
\begin{eqnarray}\label{zfactor}
  \lim_{N\rightarrow\infty } |\langle GS_{N+1} | \hat{a}_{\vec{k}_F}^\dagger |GS_N\rangle |^2 = Z_{\vec{k}_F} \neq 0 ,
\end{eqnarray}   
where $\vec{k}_F$ is an arbitrary Fermi momentum. That is, while for a Fermi liquid, we no longer have
\begin{eqnarray}
  |\langle GS_{N+1} | \hat{a}_{\vec{k}_F}^\dagger |GS_N\rangle |^2  = 1 
\end{eqnarray}
as in a Fermi gas, this quantity is still finite even in the thermodynamical limit.  Now by Equations (\ref{oexp})-(\ref{sumrule}), especially the sum rule of (\ref{sumrule}), we see that this cannot be true: At most a finite number of wave vectors can have the property that $\mathcal{O}$ is above any finite threshold. Even for this to be true, \emph{it is necessary that the unnormalized Dyson orbital of a Fermi liquid state has a finite norm even in the thermodynamical limit}. 

Finally, we also note that while the discussion above refers to fermions, the problem (\ref{defo}) and most derivations above apply also to bosons. The only modification required pertains to the upper bound of $\mathcal{O}$. Instead of Eq.~(\ref{upperb}), we have 
\begin{eqnarray}\label{upperb2}
  \mathcal{O} (f) &\equiv & |\langle GS_{N+1} | \hat{a}_{f}^\dagger |GS_N\rangle |^2 \leq  \langle GS_{N} | \hat{a}_f \hat{a}_{f}^\dagger |GS_N\rangle  \nonumber  \\
   &=&  \langle GS_{N} | 1+  \hat{a}_{f}^\dagger \hat{a}_f |GS_N\rangle   \nonumber  \\
   &\leq &  N+1 . 
\end{eqnarray}
The upper bound is achieved only when $|GS_N\rangle$ is an $N$-particle Bose-Einstein condensate occupying the single-particle orbital $f$.

\section{Laughlin wave function (fermionic case)}

The Laughlin wave function \cite{laughlin} for $N $ fermions on a plane, in the dimensionless form,  is ($z_i = x_i + i y_i $)
\begin{eqnarray}\label{laughwf}
  \Psi_{(m,N)}(z_1, \ldots, z_N ) &=& D_{(m,N)} \prod_{k=1}^{N } e^{-|z_k|^2/4} \prod_{1\leq i< j \leq N } (z_i - z_j )^m ,
\end{eqnarray}
where $m $ is an odd integer (its reciprocal is the filling factor of the fractional quantum Hall system), and $D_{(m,N)}$ is a normalization factor so that 
\begin{eqnarray}
 \int   \prod_{k=1}^N \frac{ d \bar{z}_k d z_k}{2 i }  | \Psi_{(m,N)}(z_1, \ldots, z_N ) |^2 &=& 1.  
\end{eqnarray} 
In the Laughlin wave function, the fermions live in the so-called lowest Landau level, an orthonormal basis of which is 
\begin{eqnarray}\label{landau}
  \varphi_n   (z) &=& \frac{1}{\sqrt{2\pi 2^{n} n!}} e^{-|z|^2/4}   z^n  , \quad  0\leq n \in \mathbb{Z} .
\end{eqnarray}
It is interesting to note that recently, a two-particle Laughlin state has been realized with cold atoms \cite{coldatom}.

Expanding the homogeneous polynomial $ \prod_{1\leq i< j \leq N } (z_i - z_j )^m  $ into monomials, we see that the highest power of each $z_i $ is $m(N-1)$. This means that only the first $d = m(N-1)+1$ Landau orbitals [$0\leq n \leq m (N-1)$] are involved in the Laughlin wave function $\Psi_{(m,N)} $. When $m=1$, we have $d=N$ and thus the Laughlin wave function in this case is just a Slater determinant or a Fermi sea. The $m=1$ case is therefore uninteresting and we shall consider $m=3$ or $5$ below.

We notice that 
\begin{eqnarray}\label{ratio}
  \frac{\Psi_{(m,N+1)}(z, z_1, \ldots ,z_{N})}{ \Psi_{(m,N)}(z_1, \ldots, z_N )}  &= & \frac{D_{(m,N+1)}}{D_{(m,N)} } e^{-|z|^2/4} \prod_{k=1}^{N } (z- z_k )^m .
\end{eqnarray}
Substituting (\ref{laughwf}) into (\ref{dyson2}), and making use of (\ref{ratio}), we get the unnormalized Dyson orbital
\begin{eqnarray}\label{dysonexp}
  \tilde{\phi}_D^{(m,N)} (z)&=& \sqrt{N+1 } \frac{D_{(m,N+1)}}{D_{(m,N)}} \int  \prod_{k=1}^N \frac{ d \bar{z}_k d z_k}{2 i } |\Psi_{(m,N)}|^2 e^{-|z|^2/4} \prod_{k=1}^{N } (z - z_k )^m\nonumber  \\
  &=& \sqrt{N+1 } \frac{D_{(m,N+1)}}{D_{(m,N)}}  e^{-|z|^2/4} z^{mN} \nonumber \\
  &\propto &  \varphi_{mN }(z)  . 
\end{eqnarray}
Here the second line is obtained because when the product $\prod_{k=1}^{N } (z - z_k )^m$ is expanded into monomials, only the term $z^{mN }$ which is independent of the $z_k$'s will survive the integration (Essentially, it is because $\int_0^{2\pi } d \theta e^{in \theta } = 2\pi \delta_{n 0}$ for $n \in \mathbb{Z}$ ). The third line is obtained by comparing the second line with (\ref{landau}). We see that the normalized Dyson orbital ${\phi}_D^{(m,N)} $ is simply the Landau orbital $\varphi_{mN}$! 

While (\ref{dysonexp}) shows that the maximal value of $\mathcal{O}(f) = |\langle \Psi_{(m,N+1)} | \hat{a}_{f}^\dagger | \Psi_{(m,N)} \rangle |^2$ is achieved by taking $f= 
\varphi_{mN }$, it is not useful for determining the maximal value as we do not have an analytic expression of the normalization factor $D_{(m,N)} $. To this end, we calculate directly the maximal overlap \cite{angular}
\begin{eqnarray}\label{overlap}
 \mathcal{O}_{max} = |\langle \Psi_{(m,N+1)} | \hat{a}_{\varphi_{mN}}^\dagger | \Psi_{(m,N)} \rangle |^2  .
\end{eqnarray}
We have to expand the Laughlin wave function in terms of the Slater determinants (Fock basis) constructed out of the Landau orbitals $\varphi_n$. The strategy is to use the fact that the Laughlin wave function is the zero-energy ground state of a parent hamiltonian consisting of the so-called Haldane pseudopential \cite{haldane} and solve it by exact diagonalization. Alternatively, one can also expand the product  \cite{scharf, grass} in (\ref{laughwf}), or make use of the so-called Jack polynomials \cite{jack, hu}.  The number of Landau orbitals involved is $d = m (N-1) +1 $, so the dimension of the many-body Hilbert space is $\mathcal{D} = C_d^N = \frac{d!}{N! (d-N)!}$. This number increases very quickly with $N$ and $m$. For example, for $m= 3$ and $N = 10$, we have already $\mathcal{D} = 13\;123\;110$. Fortunately, we can take advantage of the angular momentum conservation and include only those Fock states with angular momentum $m N(N-1)/2$. The dimension $\mathcal{D}_{eff}$ of the effective Hilbert space can be reduced by one or two orders. For example, $\mathcal{D}_{eff} = 246\;448$ for $(m,N)=(3,10)$.  Once we have solved the Laughlin wave functions, it is easy to calculate the overlap (\ref{overlap}). 

\begin{figure*}[tb]
\includegraphics[width= 0.475\textwidth ]{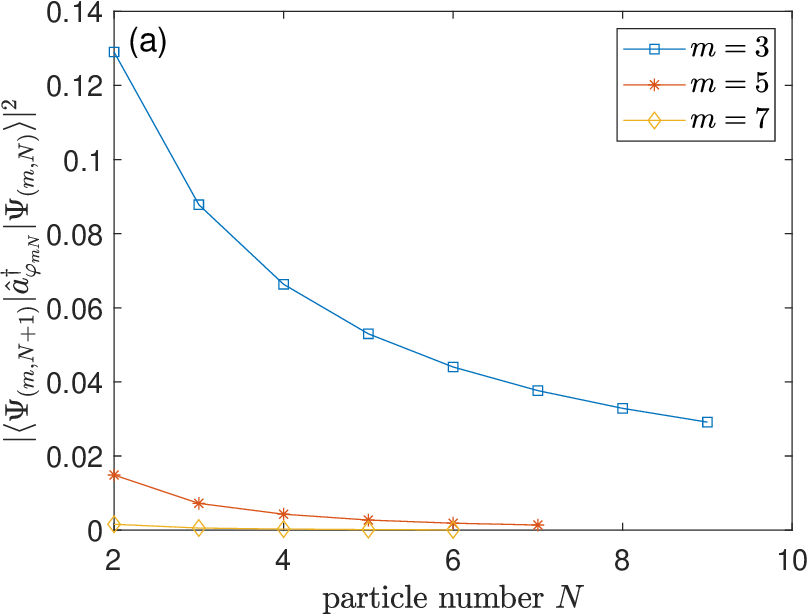}
\includegraphics[width= 0.46\textwidth ]{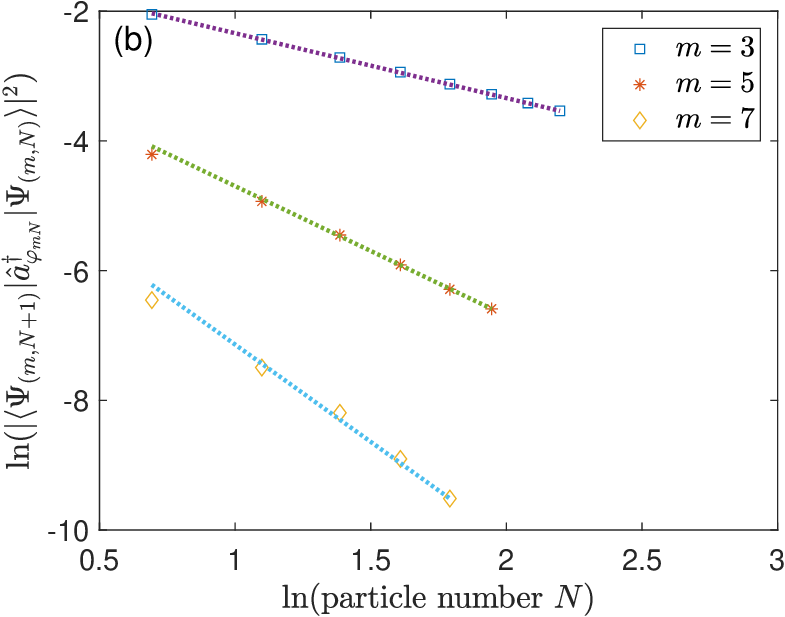}
%\hspace{11mm}
%\includegraphics[width= 0.41\textwidth ]{survival.eps}
\caption{ Overlap $ |\langle \Psi_{(m,N+1)} | \hat{a}_{\varphi_{mN}}^\dagger | \Psi_{(m,N)} \rangle |^2 $, which is the maximum of $|\langle \Psi_{(m,N+1)} | \hat{a}_{f}^\dagger | \Psi_{(m,N)} \rangle |^2$,  as a function of the particle number $N $ and for different filling factors $1/m$. Both panels display the same data, but the right one employs \emph{logarithmic} scales. The dotted straight lines there are guides for the eye. They are of slope $-1$, $-2$, $-3$ for $m=3$, $5$, $7$, respectively, indicating power-law decays of the overlap with respect to the particle number $N $. }
\label{fig1}
\end{figure*}

The results are presented in Fig.~\ref{fig1} for $m =3,5,7 $. There, we see that the value of $$|\langle \Psi_{(m,N+1)} | \hat{a}_{\varphi_{mN}}^\dagger | \Psi_{(m,N)} \rangle |^2 $$ is significantly smaller than unity,  typically by one or two orders of magnitude, even for particle numbers as small as $N\simeq 3 $, and it decreases monotonically with the electron number $N $ \cite{comment1}. This is in strong contrast to the Fermi sea case. We note that the Laughlin state $\Psi_{(m,N)} $ involves only the Landau orbitals $\varphi_n$ with $0\leq n \leq m (N-1)$, so the Dyson orbital ${\phi}_D^{(m,N)} = \varphi_{mN }$ is really a state atop and the $(N+1)$-particle state $\hat{a}_{\varphi_{mN}}^\dagger | \Psi_{(m,N)} \rangle $ is of norm unity. The small overlap of it with the Laughlin state $\Psi_{(m,N+1)}  $ indicates that there is significant relaxation or reconstruction after the introduction of the new particle. 

In Fig.~\ref{fig1}(b), we plot the same data as in Fig.~\ref{fig1}(a), but using logarithmic scales. It is interesting that for $m=3,5,7$, the data points fall on straight lines of slope $-1,-2,-3$, respectively, indicating power-law decay of the maximal overlap $\mathcal{O}_{max}$ with simple exponents. This monotonic dependence of the slope on the parameter $m$ suggests that electron-electron correlation strengthens with increasing $m$. As we shall see in next section, in the bosonic case ($m=2,4,6$), similar power-law behavior holds, and the exponent can be fitted universally with the linear function $-(m-1)/2$. 
At the time being, we do not understand these simple exponents---It should be a worthy open problem. But anyway, \emph{the power-law decay of $\mathcal{O}_{max}$ means that the $Z$ factor in (\ref{zfactor}) must be zero, and the Laughlin wave function is not a Fermi liquid. }

\begin{figure*}[tb]
\includegraphics[width= 0.5\textwidth ]{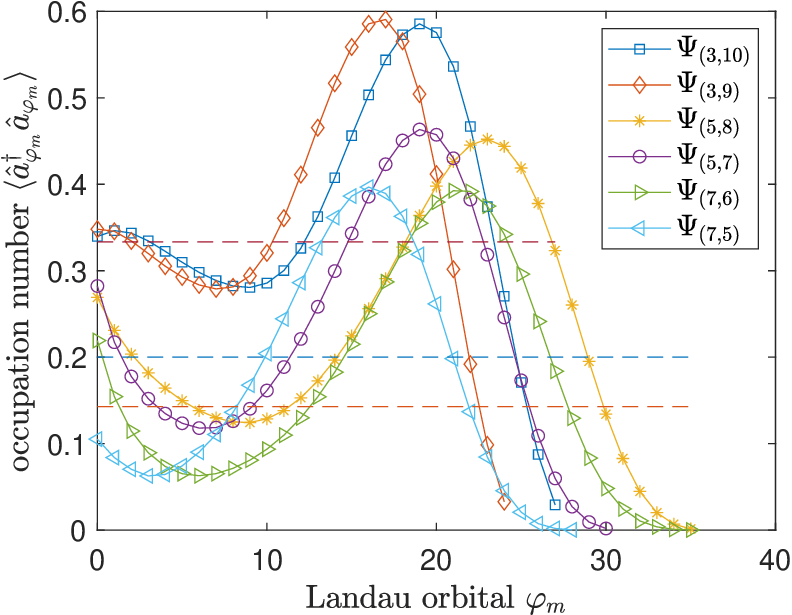}
%\hspace{11mm}
%\includegraphics[width= 0.41\textwidth ]{survival.eps}
\caption{ Occupation numbers of the Landau orbitals (\ref{landau}) for some Laughlin wave functions $\Psi_{(m,N )}$. The three dashed horizontal lines indicate the values of $1/3$, $1/5$, and $1/7$. }
\label{fig2}
\end{figure*}

\begin{figure*}[tb]
\includegraphics[width= 0.5\textwidth ]{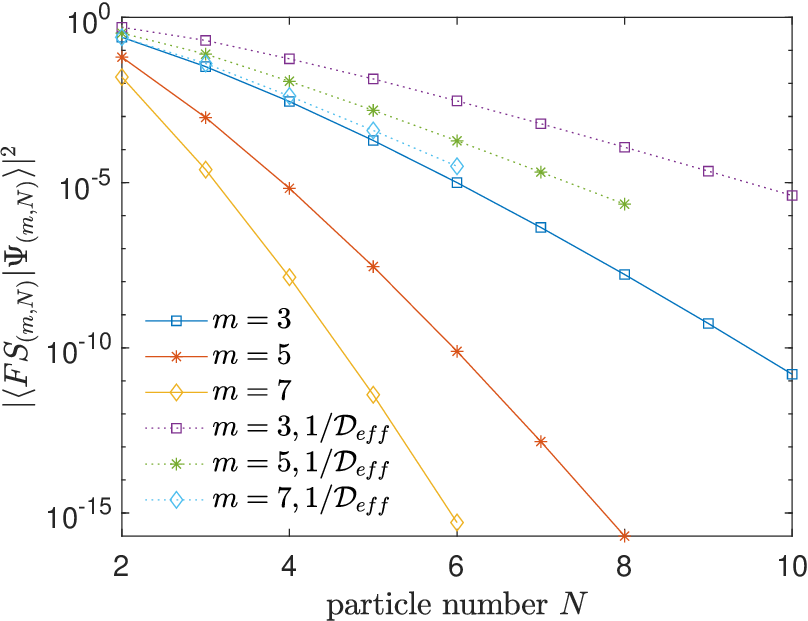}
%\hspace{11mm}
%\includegraphics[width= 0.41\textwidth ]{survival.eps}
\caption{  Overlap between the Laughlin wave function $\Psi_{(m,N)}$ and the fictitious Fermi sea $| FS_{(m,N)}\rangle $, which is constructed out of the successive Dyson orbitals as in (\ref{fermisea}). For comparison, the dotted lines indicate the value of $1/\mathcal{D}_{eff}$. }
\label{fig3}
\end{figure*}

In a Fermi sea, the Dyson orbitals have occupation number of unity and a particle placed in a Dyson orbital remains there. This is far from the case for the Laughlin wave functions. In Fig.~\ref{fig2}, we study the occupation numbers of the Landau orbitals for some Laughlin wave functions. We see smooth curves. In the Laughlin state $\Psi_{(m,N+1 )}$, the Dyson orbitals $\varphi_{mj}$ ($0\leq j \leq N$) are not more populated than other orbitals. Actually, the population on the last Dyson orbital $\varphi_{mN}$ is even a global minimum on each curve, with a value significantly smaller than unity. Note that the Landau orbital $\varphi_{mN}$ is not occupied at all in the Laughlin wave function $\Psi_{(m,N)}$, and therefore, the intermediate state $\hat{a}_{\varphi_{mN}}^\dagger |\Psi_{(m,N)}\rangle  $ is normalized and the occupancy of the orbital $\varphi_{mN}$ is unity. However, Fig.~\ref{fig2} shows that in the final state $\Psi_{(m,N+1 )}$, the occupancy of $\varphi_{mN}$ is negligibly small. The picture is then that, as we put a particle in the Dyson orbital $\varphi_{mN}$ on top of the Laughlin state $\Psi_{(m,N)}$, violent reconfigurations ensue and the population on the Dyson orbital fragments into other adjacent Landau orbitals. 

Here the more familiar concept of natural orbital \cite{lowdin} provides also some insight from another perspective. Note that by counting the angular momentum, we get easily that $\langle \Psi_{(m,N)} | \hat{a}_{\varphi_{j_1}}^\dagger  \hat{a}_{\varphi_{j_2}}| \Psi_{(m,N)} \rangle$  vanishes identically whenever $j_1  \neq j_2 $. Therefore, the single-particle reduced density matrix of a Laughlin wave function is diagonal in the Landau orbital basis, and the Landau orbitals are the natural orbitals. Now for a Fermi sea with $N$ particles, its defining feature is that it has exactly $N $ populated natural orbitals and each population is exactly unity. In contrast, here for the Laughlin wave function, we see in Fig.~\ref{fig2} that there are about $m$-fold more [$m(N-1)+1$, precisely] populated natural orbitals and the populations are around $1/m\ll  1 $. This broad and even distribution is a reflection of the strong correlation in the Laughlin wave function. 

%We see that for a Laughlin wave function $\Psi_{(m,N)}  $,  while the occupation number does peak at the Dyson orbitals $\varphi_{mk }$ ($0\leq k \leq N-1 $), Landau orbitals other than the Dyson orbitals  are also significantly populated. Roughly speaking, all the involved $d = m(N-1) +1$ orbitals have occupation numbers around the value of $1/m$. In Fig.~\ref{fig2}, for either $m$, we have examined two successive Laughlin wave functions. Comparison between them indicates how the population on the Dyson orbital $\varphi_{mN}$ fragments into other adjacent Landau orbitals. 

A Fermi sea can be completely constructed from the successive Dyson orbitals. This is again far from the case for the Laughlin state. In Fig.~\ref{fig3}, we consider the overlap between the Laughlin state $\Psi_{(m,N)}$ and the \emph{fictitious} Fermi sea state 
\begin{eqnarray}\label{fermisea}
 | FS_{(m,N)}  \rangle &=& \prod_{k=0}^{N-1} \hat{a}_{\varphi_{mk}}^\dagger |vac \rangle ,
\end{eqnarray}
which is constructed out of the successive Dyson orbitals. We see that the overlap decays with $N $ at least exponentially. It is worth emphasizing that the overlap is actually orders of magnitude smaller than $1/\mathcal{D}_{eff}$, which is the expectation value of the overlap between two randomly chosen vectors form the effective Hilbert space. 

In passing, we note that the fictitious state here corresponds to a so-called root configuration in the literature \cite{jack,hu}. 

\section{Laughlin wave function (bosonic case)}

\begin{figure*}[tb]
\includegraphics[width= 0.475\textwidth ]{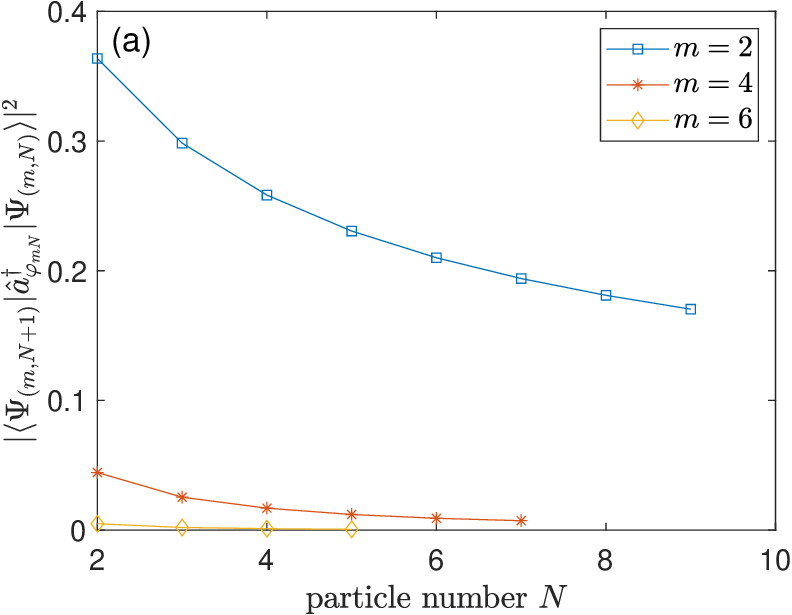}
\includegraphics[width= 0.46\textwidth ]{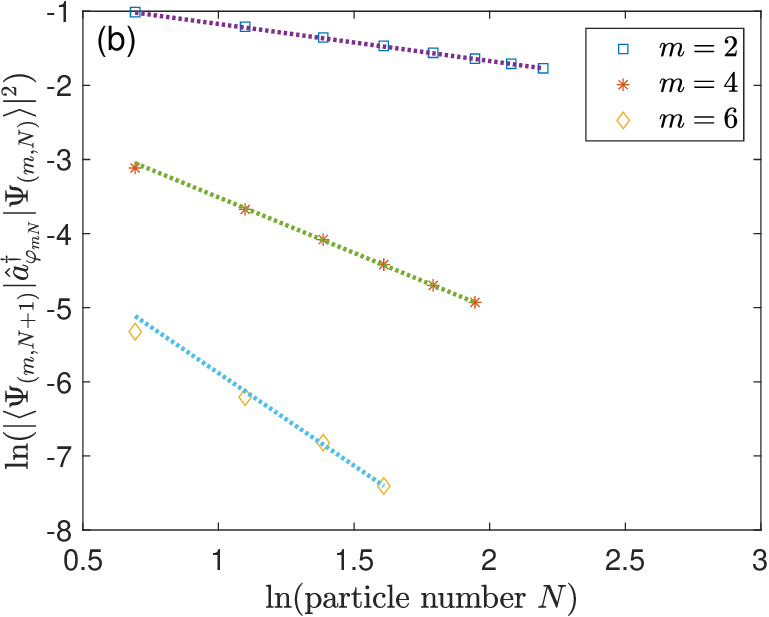}
%\hspace{11mm}
%\includegraphics[width= 0.41\textwidth ]{survival.eps}
\caption{(Similar to Fig.~\ref{fig1}, but for the bosonic case)   Overlap $ |\langle \Psi_{(m,N+1)} | \hat{a}_{\varphi_{mN}}^\dagger | \Psi_{(m,N)} \rangle |^2 $, which is the maximum of $|\langle \Psi_{(m,N+1)} | \hat{a}_{f}^\dagger | \Psi_{(m,N)} \rangle |^2$,  as a function of the particle number $N $ and for different filling factors $1/m$. Both panels display the same data, but the right one employs \emph{logarithmic} scales. The dotted straight lines there are guides for the eye. They are of slope $-1/2$, $-3/2$, $-5/2$ for $m=2$, $4$, $6$, respectively, indicating power-law decays. }
\label{fig1_bose}
\end{figure*}

\begin{figure*}[tb]
\includegraphics[width= 0.5\textwidth ]{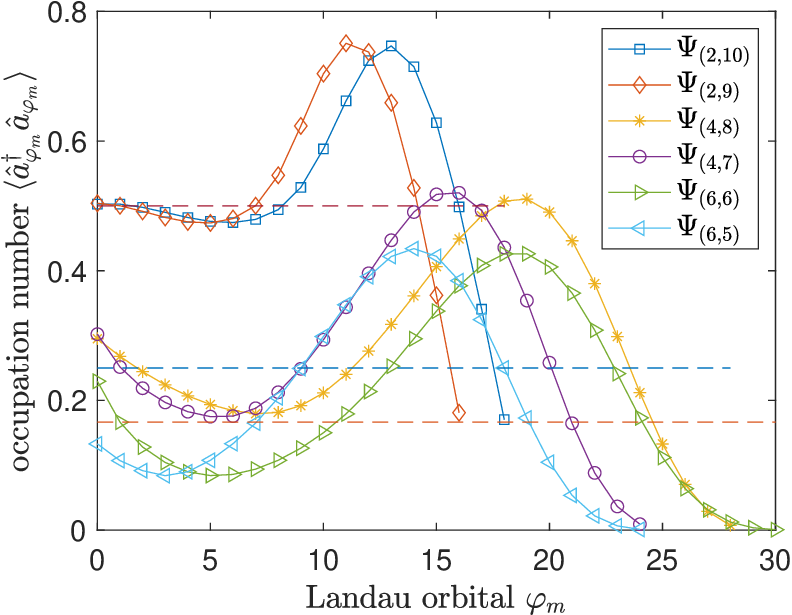}
%\hspace{11mm}
%\includegraphics[width= 0.41\textwidth ]{survival.eps}
\caption{(Similar to Fig.~\ref{fig2}, but for the bosonic case)  Occupation numbers of the Landau orbitals (\ref{landau}) for some Laughlin wave functions $\Psi_{(m,N )}$. The two dashed horizontal lines indicate the values of $1/2$, $1/4$, and $1/6$. }
\label{fig2_bose}
\end{figure*}

\begin{figure*}[tb]
\includegraphics[width= 0.5\textwidth ]{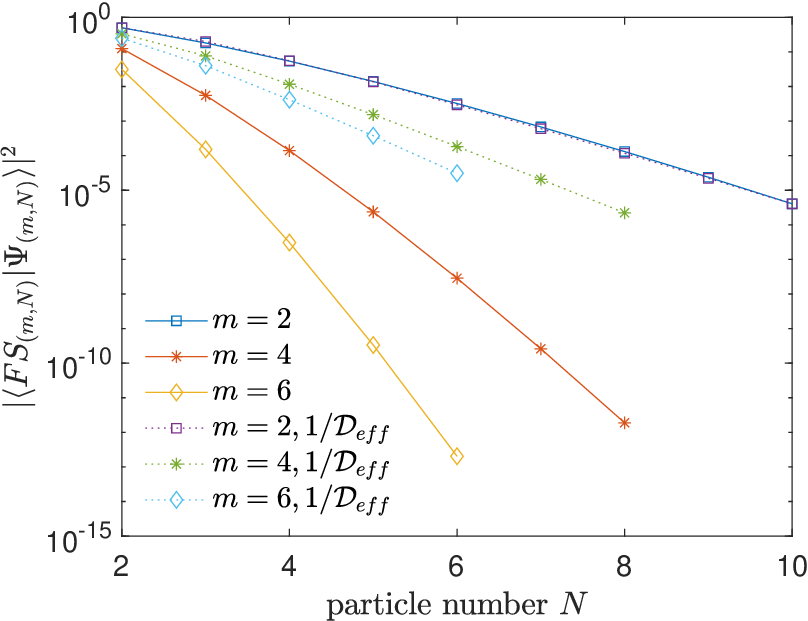}
%\hspace{11mm}
%\includegraphics[width= 0.41\textwidth ]{survival.eps}
\caption{(Similar to Fig.~\ref{fig3}, but for the bosonic case) Overlap between the Laughlin wave function $\Psi_{(m,N)}$ and the fictitious ``Fermi sea'' $| FS_{(m,N)}\rangle $, which is constructed out of the successive Dyson orbitals as in (\ref{fermisea}). For comparison, the dotted lines indicate the value of $1/\mathcal{D}_{eff}$. Note that for $m=2$, the solid line and the dotted line are very close (but not identical). }
\label{fig3_bose}
\end{figure*}

The Laughlin wave function for bosons still takes the form of (\ref{laughwf}), but with $m$ being an even integer. All the derivations in the proceeding section hold for bosons too, and the Dyson orbital is again the Landau orbital $\varphi_{mN }$. We have thus redone the three figures above, and have correspondingly Figs.~\ref{fig1_bose}-\ref{fig3_bose}. Although technically in exact diagonalization, the  fermionic and bosonic statistics have to be handled separately, and we have two independent sets of codes, the results are similar. We see similar trends and patterns in the fermionic and bosonic cases, indicating that the Laughlin wave function is a unifying framework, many properties of which are insensitive to the exchange statistics. We hypothesize that this insignificance of statistics in the Laughlin wave function arises from its low filling factor ($1/2$ at most).

Two observations are noteworthy. First, in Fig.~\ref{fig1}(b), the slopes are $-1, -2, -3$ when $m$ is $3,5,7$, respectively.  In Fig.~\ref{fig1_bose}(b), the slopes are $-1/2,-3/2,-5/2$ when $m$ is $2,4,6$, respectively. We note also that in the special case of $m=1$, the slope should be $0$ as $\mathcal{O}_{max} \equiv 1 $. In all cases, we have the power law  $\mathcal{O}_{max} \propto N^{\beta }$, with $\beta =- (m-1)/2$. This remarkably simple and universal relationship presents an intriguing open problem. 

Second, in Fig.~\ref{fig3_bose}, for $m=2$, the solid lines and the dotted lines are very close to each other (but not identical). This means that in this particular case, the root configuration has a weight close to $1/\mathcal{D}_{eff}$ in the Laughlin state. It is an intriguing question whether this is an coincidence or there is some combinatoric reason behind.

\section{Conclusions and discussions}

In conclusion, we have introduced the Dyson orbital via an optimization problem. As optimization problems are often important problems, we posit that the Dyson orbital is a conceptually and practically useful tool for characterizing many-body wave functions. It provides complementary physical insights beyond conventional metrics such as correlation functions and entanglement entropy. We acknowledge, however, that its evaluation requires priori exact knowledge of the many-body wave functions, which can be a prohibitive computational hurdle for large systems. 

We then studied the complexity of the Laughlin wave function from the Dyson-orbital perspective. Generally, the Dyson orbital cannot be calculated analytically. However, for the Laughlin wave function, because of its special structure, the Dyson orbital can be determined quite easily, although the matrix element of the associated creation operator ($\mathcal{O}_{max}$) is still laborious to obtain. We find that each time the particle number increases by one, the wave function undergoes significant reconfiguration, which shows that it is indeed incompatible with the picture of filling single-particle orbitals one by one. That is, it is not a Fermi gas or even a Fermi liquid.  While this might be well known or well accepted, our data at least has shed new light on the structure of the Laughlin wave function. \emph{Quantitatively}, we find that $\mathcal{O}_{max}$ decays with the particle number according to a \emph{power law} with very simple exponents (Fig.~\ref{fig1} and Fig.~\ref{fig1_bose}), which unfortunately, we cannot explain at the time being. But anyway, the mere power-law decay \emph{proves quantitatively} that the Laughlin wave function corresponds to a non-Fermi liquid. It is also somewhat unexpected that the decay follows a power law, rather than an exponential one, in view of the folklore that the Laughlin wave function is a strongly correlated system.

The present study of the Laughlin wave function establishes a proof-of-principle application of the Dyson orbital method. More systematic studies of many-body wave functions in various models from the Dyson orbital perspective are warranted. The fractional quantum Hall effect is but one instance of strongly interacting non-Fermi liquids. Future work will extend this analysis to other canonical non-Fermi liquids, such as Luttinger liquids \cite{giuliani} and the enigmatic strange metals \cite{strange}. The Sachdev-Ye-Kitaev (SYK) \cite{syk1,syk2} model presents a particularly attractive starting point, owing to its explicit Hamiltonian and the fact that its ground state can be solved numerically with relative ease. This tractability will allow for a direct and rigorous analysis using the Dyson orbital framework. 

Valuable insights might be gained by investigating how $\mathcal{O}_{max}$ varies with the particle number $N $. For example, closed shells might be confirmed or revealed as local maxima of $\mathcal{O}_{max}$, because of their resilience to particle addition \cite{dot}.  The asymptotic behavior of $\mathcal{O}_{max}$ is also of great interest. We note that our study concerns the overlap of two many-body wave functions, a concept also relevant to the orthogonality catastrophe \cite{anderson}, where the power-law exponent is connected to phase shifts.

Finally, we think our work is of pedagogical value, as it makes the Laughlin wave function useful for introducing the notions of Dyson orbital and non-Fermi liquid.

%We thus think the Laughlin wave function is useful for introducing the notions of Dyson orbital and non-Fermi liquid, and the calculations in this paper can sever as exercises in undergraduate or graduate courses on quantum mechanics or solid state physics. 

\section*{Acknowledgments}

The authors are grateful to J. Guo and R. Huang for their helpful comments. This work is funded by National Key R\&D Program of China under Grant No.
2021YFB3501503. 

% National Natural Science Foundation of China (Grant Nos. U2230401） 
% Special Fund for Research on National Major Research Instrument (No. 62327804) and the Foundation of LCP

%\section*{References}


\begin{thebibliography}{99}
\bibitem{phe}
Here we consider the particle excitation, which involves an $N$-particle state and an $(N+1)$-particle state. The problem can be generalized, say, to the particle-hole excitation scenario. What single-particle (normalized) orbitals $f_{1,2}$ can maximize the transition amplitude $|\langle \Psi_N^{(2)}|\hat{a}_{f_1}^\dagger \hat{a}_{f_2} | \Psi_N^{(1)}\rangle |$, where $|\Psi_N^{(1,2)}\rangle $ are two given $N$-particle states?
\bibitem{angular}
We note in passing that by counting the total angular momentum, it is easy to see that $\langle \Psi_{(m,N+1)} | \hat{a}_{\varphi_{j}}^\dagger | \Psi_{(m,N)} \rangle $ vanishes identically for $j \neq m N $.
\bibitem{comment1}
At least for the Laughlin wave function which describes a droplet with a size increasing with the electron number, the monotonic decrease of $\mathcal{O}_{max}$ is expected, as more electrons means more degrees of freedom. However,  non-monotonicity of $O_{max}$ should be possible and can be of physical significance.
\end{thebibliography}
\end{document}